\documentclass[aps,pra,twocolumn,superscriptaddress]{revtex4-1}
\usepackage[utf8]{inputenc}
\usepackage{braket,textcomp}

\usepackage{color}
\usepackage{amssymb,amsmath,bbm,mathcomp}
\usepackage{times}

\newcommand{\smt}{{\!\scriptscriptstyle T\,}}

\newcommand{\sms}{{\!\scriptscriptstyle S\,}}
\newcommand{\ou}{\text{\tiny OU}}

\newcommand{\exval}[1]{\langle{#1}\rangle}

\DeclareMathOperator{\sign}{sgn}

\usepackage[breaklinks, colorlinks=true, urlcolor=blue, anchorcolor=blue, citecolor=blue, filecolor=blue, linkcolor=blue, menucolor=blue, linktocpage=true, pdfproducer=medialab, pdfa=true]{hyperref}

\begin{document}
\title{Effective description of the short-time dynamics in open quantum systems}
\author{Matteo A. C. Rossi}
\email{matteo.rossi@unimi.it}
\affiliation{Quantum Technology Lab, Dipartimento di Fisica, Università
degli Studi di Milano, I-20133 Milano, Italy}
\author{Caterina Foti}
\affiliation{Dipartimento di Fisica e Astronomia, Universit\`a di Firenze, I-50019, Sesto Fiorentino (FI), Italy}
\affiliation{INFN, Sezione di Firenze, I-50019, Sesto Fiorentino (FI), Italy}
\author{Alessandro Cuccoli}
\affiliation{Dipartimento di Fisica e Astronomia, Universit\`a di Firenze, I-50019, Sesto Fiorentino (FI), Italy}
\affiliation{INFN, Sezione di Firenze, I-50019, Sesto Fiorentino (FI), Italy}
\author{Jacopo Trapani}
\affiliation{Quantum Technology Lab, Dipartimento di Fisica, Università
degli Studi di Milano, I-20133 Milano, Italy}
\author{Paola Verrucchi}
\affiliation{ISC-CNR, UOS Dipartimento di Fisica, Universit\`a di Firenze, I-50019, Sesto Fiorentino (FI), Italy}
\affiliation{Dipartimento di Fisica e Astronomia, Universit\`a di Firenze, I-50019, Sesto Fiorentino (FI), Italy}
\affiliation{INFN, Sezione di Firenze, I-50019, Sesto Fiorentino (FI), Italy}
\author{Matteo G. A. Paris}
\affiliation{Quantum Technology Lab, Dipartimento di Fisica, Università
degli Studi di Milano, I-20133 Milano, Italy}
\affiliation{INFN, Sezione di Milano, I-20133, Milano, Italy}
\date{\today}
\begin{abstract} We address the dynamics of a bosonic system coupled to either a
bosonic or a magnetic environment, and derive a set of sufficient conditions
that allow one to describe the dynamics in terms of the effective interaction
with a classical fluctuating field. We find that for short interaction times the
dynamics of the open system is described by a Gaussian noise map for several
different interaction models and independently on the temperature of the
environment. In order to go beyond a qualitative understanding of the
origin and physical meaning of the above short-time constraint, we take a
general viewpoint and, based on an algebraic approach, suggest that any quantum
environment can be described by classical fields whenever global symmetries lead
to the definition of environmental operators that remain well defined when
increasing the size, i.e. the number of dynamical variables, of the environment.
In the case of the bosonic environment this statement is exactly demonstrated
via a constructive procedure that explicitly shows why a large number of
environmental dynamical variables and, necessarily, global symmetries, entail
the set of conditions derived in the first part of the work.
\end{abstract}
\maketitle
\section{Introduction}
The modeling of any open quantum system (OQS) inherently implies that of its
surroundings. However, knowing the quantum structure of the total
Hamiltonian, including the details of the couplings between the
principal system A and its environment $\Xi$, does not usually
suffice to develop a simple and meaningful model of the overall system,
due to $\Xi$ being made of a very large number $N$ of quantum components,
a fact that we will hereafter take as integral to the definition of
{\em environment}. On the other hand, knowing specific features of $\Xi$
may help selecting a suitable formalism and/or some appropriate
approximations, so as to devise the most effective strategies for
tackling problems that cannot be otherwise studied.
\par
As a matter of fact, the modeling of an effective description of
$\Xi$ and of its influence on A usually stems from intuitive and
phenomenological arguments \cite{alicki07},
or even from an arbitrary choice, rather than a formal derivation.
One of the reasons why this is so typical in the study of OQS is that the large-$N$ theories that have been extensively developed
and used in quantum-field-theory since the 1970s (comprehensive
bibliographies and discussions can be found for instance in
Refs.~\cite{BrezinW93,Auerbach94}) are not
trivially applicable when the large-$N$ system is not isolated, but
rather coupled
with a small, invariably quantum, principal system. Unless one decides
that the latter is not ``principal'' at all, and can be hence neglected,
several foundational issues arise in this setting, due to the difficult
coexistence of quantum and classical formalisms, possibly made worse by
the presence of thermal baths or stochastic agents.
\par
Having this issue in mind, here we analyze a specific
situation where a principal quantum system A interacts with an equally
quantum environment $\Xi$, which is put into contact with a further
external system T. If $\Xi$ is macroscopic and T is a thermal bath at
high temperature, it may appear intuitive, and naively
understood, that A effectively evolves as if it were under the influence
of a classical fluctuating field. This statement, however, has the nature
of an ansatz as far as it is not formally inferred, and conditions
ensuring its validity are not given.
\par
Several OQS have been indeed
investigated to assess whether an effective
description is viable,  where the effects of the environment are described either with a stochastic Schrödinger equation \cite{str99,sto02,lacroix08} or in terms of the coupling with a classical fluctuating field
\cite{Helm2009,Helm2011,Crow2014,witzel14,Palma567,%
Paladino2002,Cywiski2008,Wold2012,Hung2013,Paladino2014,%
LoFranco2016,Orieux2015,LoFranco2012,Leggio2015,Xu2013}.
In the latter case, full equivalence has been shown only
for single-qubit dephasing dynamics \cite{Crow2014}, with an explicit
construction of the corresponding classical stochastic
process. General arguments valid also for bipartite systems
have been discussed \cite{Benedetti2014d,Shao04,Rossi2014b}
and the effects of the interaction with a classical field have
been investigated in detail
\cite{Trapani2016,Galperin2006,Benedetti2013a,Yu2010,Benedetti2014c,Rossi2016,
Benedetti2014b,Rossi2015}. Parametric representation have also been used
to show that classical variables can emerge in quantum Hamiltonians as
environmental degrees of freedom
\cite{CalvaniCGV13,Liuzzo-ScorpoCV15b,FotiCV16,Strunz05,PerniceHS12,PerniceS11}.
 \par
In this work we scrutinize the general idea
that the dynamics of a quantum system with a macroscopic
environment may be effectively described by a non-autonomous, i.e.
time-dependent, Hamiltonian acting on the principal system only.
In particular, we critically inspect the conditions for the validity
of this hypothesis as a tool to understand whether it stems from
$\Xi$ being macroscopic, or the temperature being high, or from
enforcing some other specific condition.
\par
To this aim, in Secs.~\ref{s.bosonic} and \ref{s.magnetic} we study
two specific models that go beyond the pure-dephasing, and whose
analysis will also serve as explicit guidance for the more abstract
approach presented in Sec.~\ref{s.large-N}.

In particular, in Sec.~ref{s.bosonic} we
consider the case where ${\rm A}$ is a bosonic mode coupled with an
equally bosonic environment, hereafter called ${\rm B}$, which is made
of $N$ distinguishable modes that do not interact amongst themselves.
The Hamiltonian
reads
  \begin{align}
  H={} & \nu a^\dagger a +\sum_k^N(\lambda_{1k} a^\dagger+\lambda_{2k}a)b_k \notag \\
  & +\sum_k^N(\lambda_{1k}^* a+\lambda_{2k}^*a^\dagger)b_k^\dagger
  + \sum_k^N \omega_k b_k^\dagger b_k
  \label{e.H-bosonic_k}
\end{align}
where
$[a,a^\dagger]=1$ and $[b_k,b^\dagger_{k'}]=\delta_{kk'}$, with
$\nu,\omega_k\in\mathbb{R}$ and $\lambda_{1k},\lambda_{2k} \in
\mathbb{C},~\forall k$. Also, we have set $\hbar=1$, as done throughout this work.
Studying the evolution of the reduced density matrix for the
principal system, we show that the short-time dynamics defined by
Eq.~\eqref{e.H-bosonic_k} with $\omega_k \simeq \omega$ $\forall k$, can be
described by an effective Hamiltonian acting on A only, $H^{\rm
eff}_{\rm A}(\zeta)$, where the functions $\zeta$ embody the remnants of
${\rm B}$ in the form of classical, possibly fluctuating fields,
depending on external parameters such as time and temperature.
In what follows we will refer to the condition $\omega_k \simeq \omega$ $\forall k$
as defining a \emph{narrow energy spectrum}.
\par
The above model has a sibling that describes the case
of a spin environment, hereafter called S, made by $N$ distinguishable
spin-$\frac{1}{2}$ particles that do not interact amongst themselves.
Its dynamics is studied in Sec.~\ref{s.magnetic}, as described by the
Hamiltonian
\begin{align}
  H^\sms= {}&\nu a^\dagger a
  +\sum_i^N(g_{1i} a^\dagger+g_{2i}a)\sigma^-_i  \notag \\
  & +\sum_i^N(g_{1i}^* a+ g_{2i}^*a^\dagger)\sigma_i^+
  + \sum_i^N f_i \sigma^z_i
\label{e.H-spin_i}
\end{align}
where
$[\sigma^+_i,\sigma^-_{i'}]=2\delta_{ii'}\sigma_i^z$,
$[\sigma^z_i,\sigma^\pm_{i'}]=\pm\delta_{ii'}\sigma^\pm_i$,
$f_i\in\mathbb{R}$ and $g_{1i},g_{2i} \in
\mathbb{C},~\forall i$.
Despite differences with the case of a bosonic environment emerge,
essentially due to the specific algebra of the spin operators, the
short-time dynamics of this model for $f_i\sim f$ $\forall i$  is also
found to be properly described by an effective Hamiltonian $H^{\rm
eff}_{\rm A}(\zeta)$.
\par
Upon inspecting the dynamics of both systems
in order to retrace the derivation of the short-time dynamics,
we notice that no explicit condition on the
value of $N$ is involved. This is somehow surprising, given that B and S are
named {\em environment} insofar as the number $N$ of their
quantum components is large, virtually infinite in the case
of a macroscopic environment.
Therefore, in order to understand whether a relation exists between a
large value of $N$ and the assumptions of
short-time and narrow energy spectrum $\omega_k \simeq \omega$ used in
Secs.~\ref{s.bosonic}-\ref{s.magnetic}, in Sec.~\ref{s.large-N} we take on
the model \eqref{e.H-bosonic_k} from a more abstract viewpoint. More specifically:
we generalize the well-established procedures for
deriving classical theories as large-$N$ limit of quantum
ones~\cite{Yaffe1982} to the case of composite quantum systems, and
find that replacing quantum operators  by classical
fields for $N\to\infty$ requires that environmental
operators stay well defined in such limit which, in turn,
implies the environment to feature some global symmetry.
In particular, we show that the renormalization of the couplings, which
is necessary for the $N\to\infty$ limit to stay physically meaningful,
reflects upon the short-time condition previously used. Also, we
will discuss how a narrow environmental energy spectrum
$\omega_k \simeq \omega$ $\forall k$ is the key feature that
guarantees the existence of a global symmetry in the theory
defined by Eq.~\eqref{e.H-bosonic_k},
namely the symmetry under permutation of different modes.
\par
Overall, collecting our diverse results, we put forward the
conjecture that non-autonomous Hamiltonians for closed quantum systems
describe the short-time dynamics of interacting models involving at
least one macroscopic subsystem. We also comment upon the symmetry
properties allowing this subsystem to emerge as a macroscopic
one, and the related features of its energy spectrum. Finally, we
discuss the role of such symmetry properties in the design of a
general procedure for deriving an effective non-autonomous Hamiltonian
from an interacting microscopic model.
\par
Sec.~\ref{s.conclusions} closes the paper with some concluding remarks.
\section{Bosonic environment}
\label{s.bosonic}

We consider the Hamiltonian \eqref{e.H-bosonic_k},
for either
1) $\lambda_{2k}=0$, with $\lambda_k\equiv\lambda_{1k}$ finite (linear exchange), or
2) $\lambda_{1k}=0$, with $\lambda_k\equiv\lambda_{2k}$ finite (parametric hopping),
$\forall k$, i.e.
\begin{align}
H_1 &= \nu a^\dag a + \sum_k \omega_k b_k^\dag b_k + \sum_k \left(
\lambda_k^* a b_k^\dag + \lambda_k a^\dag b_k \right)
,\label{e.bosonic-ex}\\
H_2 &= \nu a^\dag a + \sum_k \omega_k b_k^\dag b_k + \sum_k \left(
\lambda_k^* a^\dag b_k^\dag + \lambda_k a b_k \right) .
\end{align}

We hereafter use the index $j=1,2$ to refer to the exchange and hopping case, respectively.
The Heisenberg equations of motion (EOM) for the mode operators are
\begin{subequations}\label{e.EOM-D}
  \begin{align}
  \text{Exchange:}\qquad \dot a &= i [H_1,a] =
  - i \nu a - i \sum_k \lambda_k b_k
  \,, \\
  \dot b_k & = i [H_1,b_k] = - i \omega_k b_k - i \lambda_k^* a \,,
  \end{align}
\end{subequations}
\vspace*{-\abovedisplayskip}
\begin{subequations}\label{e.EOM-A}
  \begin{align}
\text{Hopping:}\qquad  \dot a &= i [H_2,a] =
- i \nu a- i \sum_k \lambda_k^* b_k^\dag
\,, \\
\dot b^\dag_k &  = i [H_2,b_k^\dag] =
 i \omega_k b^\dag _k  + i \lambda_k a \,.
\end{align}
\end{subequations}

If the spectrum of the environment is narrow enough to write
$\omega_k\simeq \omega~\forall k$,
the above EOM can be written as
\begin{align}
&\text{Exchange:}~~~~\dot a   = - i \nu a - i \Lambda b~~\,,~
\dot b = - i \omega b - i \Lambda a\,,\label{e.gl-EOM-D}\\
&\text{Hopping:}~~~~\dot a   = - i \nu a - i \Lambda b^\dag~\,,~
\dot b^\dag   = i \omega b^\dag + i \Lambda a\,,
\label{e.gl-EOM-A}
\end{align}
where the bosonic operator $b$
is defined as
\begin{equation}
b \equiv\frac{1}{\Lambda}\sum_k\lambda_kb_k\,,
\label{e.global-op-unnorm}
~~~{\rm with}~~~
\Lambda^2\equiv\sum_k|\lambda_{k}|^2\,.
\end{equation}
The above Eqs.~(\ref{e.gl-EOM-D}-\ref{e.gl-EOM-A})
are the same EOM that one would obtain starting from the two-mode
bosonic Hamiltonians
\begin{align}
&\hbox{Exchange:}~~~~
\nu a^\dagger a+\omega b^\dagger b+ \Lambda (a b^\dagger+a^\dagger b)\,,
\label{e.global-HD}\\
&\hbox{Hopping:}~~~~
\nu a^\dagger a+\omega b^\dagger b+ \Lambda (a^\dagger b^\dagger+a b)\,,
\label{e.global-HA}
\end{align}
describing two oscillators, with different frequencies $\nu$ and
$\omega$, exchanging quanta through a linear interaction.
Notice, though, that such direct relation only exists in the case of a
narrow spectrum, $\omega_k\sim\omega$, $\forall k$.

Both systems of Eqs. \eqref{e.gl-EOM-D} and \eqref{e.gl-EOM-A} can
be solved by Laplace transform, using the rule $\tilde{\dot a} (s) = s
\tilde a
(s) - a(0)$ to obtain algebraic equations from differential ones.
Few calculations lead us, after back-transforming and recalling that the
index $j=1,2$ refers to the Exchange,Hopping respectively, to the
solutions
\begin{align}
a(t) & = e^{-i H_j t}\, a\, e^{i H_j t} =
\left[\mu_j(t)\, a + \pi_j(t)\, {\cal B}_j \right]\,
e^{- i \omega_j t}\,, \notag \\
{\cal B}_j(t) & = e^{-i H_j t}\, {\cal B}_j\, e^{i H_j t} =
\left[(-)^j\pi_j^*(t)\, a + \mu_j^*(t)\, {\cal B}_j \right]\,
e^{- i \omega_j t}\,,
\label{e.EOM-solutions}
\end{align}
where ${\cal B}_{1}=b$, ${\cal B}_{2}=b^{\dag}$,
\begin{subequations}
\label{e.mu-pi_t}
\begin{align}
\mu_j(t) &=\cos(\Delta_j t) -i \frac{\delta_j}{\Delta_j} \sin(\Delta_j t)~\,,\label{e.mu_t}\\
\pi_j(t) &= -i\frac{\Lambda}{\Delta_j} \sin (\Delta_j t)\,,
\label{e.pi_t}
\end{align}
\end{subequations}
with
\begin{subequations}
\label{e.dDelta_j}
\begin{align}
\delta_j &=\frac{1}{2}\left(\nu+(-)^j\omega\right)~\,, \\
\omega_j &=\frac{1}{2}\left(\nu-(-)^j\omega\right)~\,, \\
\Delta_j^2&=|\delta_j^2-(-)^j\Lambda^2|~\,,
\end{align}
\end{subequations}
and we have used $\mu^*_j(t)=\mu_j(-t)$.
The overall phase factors in the rightmost terms of Eqs.~\eqref{e.EOM-solutions} suggest that
a natural interaction picture exists,
corresponding to frames rotating at frequency $\omega_j$.
We will use these frames in the following, so as to omit those phase factors.
Further notice that $|\mu_j(t)|^2-(-)^j |\pi_j(t)|^2 =1$, ensuring that
$[a(t), a^\dag (t)]=[b(t),b^\dag (t)]=1$, $\forall t$ and also that
$|\mu_j(t)|^2 +(-)^j \pi^2_j(t) =1$, meaning that the evolutions
correspond to rotations in the rotating frames.

Our goal is now to obtain an effective Hamiltonian
$H_{\rm A}^{\rm eff}(\zeta)$, acting on A only, without renouncing to the quantum character
of its companion B. This means that we can consider nothing but the time
dependence of the reduced density matrix for A
\begin{equation}
\rho_{\rm A}(t)=
\hbox{Tr}_{\rm B}\left[ e^{-i H_j t}
\rho_{\rm A}\otimes\rho_{\rm B}\, e^{i H_j t} \right]
\equiv {\cal E}_j[\rho_{\rm A}](t)\,,
\label{e.rho-A(t)}
\end{equation}
with the notation $\rho_{\rm X}\equiv\rho_{\rm X}(0)$ used hereafter.
In particular, as already implied by Eq.~\eqref{e.rho-A(t)}, we want to derive the explicit form of the
dynamical map ${\cal E}_j[\rho_{\rm A}]$ upon assuming that at $t=0$ the system A+B is in a
factorized state $\rho_{\rm A}\otimes\rho_{\rm B}$. Moreover, we specifically take B initially
prepared in the state at thermal equilibrium
\begin{equation}
\rho_{\rm B}= \frac{1}{1+n_\smt}
\left(\frac{n_\smt}{1+n_\smt}\right)^{b^\dag b}\,,
\label{e.rhoB-0}
\end{equation}
where $n_\smt=(e^{\omega/T}-1)^{-1}$ is the thermal number of
photons, and we have set the Boltzmann constant equal to 1.

After this choice, that implicitly means that B further interacts
with a third system T, specifically a thermal bath due to the choice of
the state in Eq. \eqref{e.rhoB-0}, we can positively move towards the
derivation of the field $\zeta$ entering $H^{\rm eff}_{\rm A}$,
and of its possible dependence on some
external parameter.
To this aim we first write the initial state of A+B using the
Glauber formula,
\begin{align}
&\rho_{\rm A} \otimes\rho_{\rm B} = \notag \\ & \iint\!\frac{d^2\gamma'd^2\gamma''}{\pi^2}
\chi[\rho_{\rm A}](\gamma ')
\chi[\rho_{\rm B}](\gamma '')
D_a^\dag (\gamma ')\otimes D_b^\dag (\gamma '')\,,
\label{e.glauber-initial}
\end{align}
where
$\chi[\rho](\gamma)=\hbox{Tr}[\rho\,D(\gamma)]$ is the
characteristic function of the state $\rho$, and
$D_x(\gamma) = \exp\{\gamma x^\dag -\gamma^* x\}$, with $[x,x^\dag]=1$,
is the bosonic displacement operator.
In order to get the argument of the partial trace in Eq.~\eqref{e.rho-A(t)},
we use Eqs.~\eqref{e.EOM-solutions} to write the evolution of the
displacement operators entering Eq.~\eqref{e.glauber-initial},
\begin{align}
e^{-i H_j t} & D_a^\dag (\gamma ')
\otimes D_b^\dag (\gamma '')\, e^{i H_j t} = \notag \\
& D_a^\dag [\mu_j^*(t) \gamma '  + \pi_j^*(t) \gamma '' ]
\otimes D_b^\dag [\pi_j^*(t) \gamma '  + \mu_j(t) \gamma '' ]\,.
\label{e.displacement-t}
\end{align}
We then perform the partial trace using
$\hbox{Tr}\left[D(\gamma)\right] = \pi \delta^{(2)}(\gamma)$, so as to
get
\begin{align}
{\cal E}_j & [\rho_{\rm A}](t) \notag \\
& = \int\!\frac{d^2\gamma '}{\pi}
\chi[\rho_{\rm A}](\gamma ')\,
\chi[\rho_{\rm B}]
\left(-\frac{\gamma '\pi^*_j(t)}{\mu_j(t)}\right)\,
D^\dag \left(\frac{\gamma '}{\mu_j(t)}\right) \notag \\
& = \int\!\frac{d^2\gamma}{\pi}
\left|\mu_j(t)\right|^2
\chi[\rho_{\rm A}](\gamma \mu_j(t))\,
\chi[\rho_{\rm B}](- \gamma \pi^*_j(t))\, D^\dag (\gamma)\,,
\label{e.dmap}
\end{align}
where, in the last step, we made the substitution $\gamma'\to\gamma\mu_j(t)$.

Upon expanding the coefficients \eqref{e.mu-pi_t} for $\Delta_j t\ll 1$,
\begin{subequations}\label{e.short-time-expansion}
  \begin{align}
  \mu_j(t) &\simeq 1 -i \delta_j t + O(t^2)~\,,\\
  \pi_j(t) &\simeq - i \Lambda t + O(t^2)~\,,\\
  |\mu_j(t)|^2 &\simeq 1 + O(t^2)\,,
  \end{align}
\end{subequations}
and using the explicit form of the characteristic function of the
thermal state,
$\chi[\rho_{\rm B}](\gamma) = \exp\{-|\gamma|^2 (n_\smt +\frac12)\}$,
we finally write
\begin{align}
\rho_{\rm A}(t) & ={\cal E}_j[\rho_{\rm A}](t) \notag\\
& = \int\!\frac{d^2\gamma}{\pi}
\chi[\rho_{\rm A}](\gamma) e^{-|\gamma|^2\sigma^2(t)}\, D^\dag
(\gamma)\,,
\label{e.mapA}
\end{align}
with $\sigma^2(t)=\Lambda^2  t^2 (n_\smt + \tfrac 12)$.

We now wonder whether the above map is realized by some known
unitary evolution involving the interaction with a
classical environment only. Indeed, by first noticing
that for any state $\varrho$ it is
\begin{equation}
\chi[\varrho](\gamma) e^{-|\gamma|^2\sigma^2}=\chi[\varrho_{_{\rm GN}}](\gamma)\,,\label{e.chign-rhogn}
\end{equation}
with
\begin{equation}
\varrho_{_{\rm GN}}\equiv\int\! \frac{d^2 \alpha}{\pi \sigma^2}
e^{-\frac{|\alpha|^2}{\sigma^2}}\, D(\alpha)\varrho\, D^\dag(\alpha)\,,
\label{e.kraus-decomposition}
\end{equation}
we recognize in Eq.~\eqref{e.kraus-decomposition} the
Kraus decomposition corresponding to a Gaussian noise (GN) channel,
namely a random displacement with Gaussian modulated amplitude
\footnote{
To prove the equivalence in Eq. \eqref{e.chign-rhogn}, start from
$\hbox{Tr}[\rho_{\rm GN}\, D(\gamma)]$,
insert the definition of $\rho_{\rm GN}$, use the composition rule $D^\dag(\alpha)D(\gamma)D(\alpha) =
D(\gamma) e^{\alpha^* \gamma - \gamma^* \alpha}$ and perform the
resulting Fourier transform.}.
\par
The same map \cite{Puri2001,Trapani2015} describes the
evolution of a bosonic system in the presence of a classical fluctuating
field, i.e. governed by a non-autonomous Hamiltonian of the form
\begin{align}
H_{\rm stoc}(t) =
\nu a^\dag a  + a \zeta^*(t) e^{i \omega_\zeta t} + a^\dag \zeta(t)
e^{- i \omega_\zeta t}\,,
\label{e.Hstoc}
\end{align}
where $\zeta(t)$ is a random classical field described by a Gaussian
stochastic process $\zeta(t)=\zeta_x(t) + i \zeta_y(t)$ with zero mean
$[\zeta_x(t)]_\zeta = [\zeta_y(t)]_\zeta=0$ and diagonal structure of the
autocorrelation function
\begin{subequations}
\label{e.kernel}
\begin{align}
\left[\zeta_x (t_1) \zeta_x (t_2)\right]_\zeta & =
\left[\zeta_y (t_1) \zeta_y (t_2)\right]_\zeta =
K(t_1,t_2)\,, \\
\left[\zeta_x (t_1) \zeta_y (t_2)\right]_\zeta & =\left[\zeta_y (t_1)
\zeta_x (t_2)\right]_\zeta =
0\,.
\end{align}
\end{subequations}
The function $\sigma(t)$ in Eq. \eqref{e.chign-rhogn} is in this case
\begin{equation}
  \sigma(t) = \int_{0}^t\int_0^t  dt_1 dt_2 \cos[\delta_\zeta(t_1-t_2)]
K(t_1,t_2)\,,
\end{equation}
where $\delta_\zeta = \omega_\zeta - \nu$ is the detuning between the natural
frequency $\nu$ of A and the central frequency $\omega_\zeta$
of the classical field $\zeta(t)$.
The map \eqref{e.chign-rhogn} may be obtained, for instance, upon
considering the classical environment fluctuating according to a Gaussian
Ornstein-Uhlenbeck stochastic process \cite{Uhlenbeck1930} characterized by the autocorrelation function
\begin{equation}
	K^\ou_\tau(t_1-t_2) = \frac{G}{2\tau }
e^{-\frac{1}{\tau}|t_1-t_2|}\,,
\end{equation}
where $\tau$ is the correlation time, and $G$ is the
amplitude of the process. In the short-time limit, one easily finds that
\begin{equation}
  \sigma(t) =  \frac{G}{2\tau} t^2\,.
\label{e.sigma}
\end{equation}

In conclusion, we have shown that, as far as $t\ll|\Delta_j|^{-1}$,
the effective Hamiltonian $H_{\rm A}^{\rm eff}(\zeta(t))$ equals $H_{\rm
stoc}(t)$, meaning that
\begin{equation}
H_{\rm A}^{\rm eff}(\zeta(t))=
\nu a^\dag a  + a \zeta^*(t) e^{i \omega_\zeta t} + a^\dag \zeta(t)
e^{- i \omega_\zeta t}\,,
\label{e.HAeff}
\end{equation}
with the field $\zeta(t)$ as from Eqs.~(\ref{e.kernel}-\ref{e.sigma}),
and $G = 2\tau{\Lambda^2}(n_\smt + \tfrac 12)$.

Notice that the dynamical map for A in the short-time limit, Eq.
(\ref{e.mapA}), is the same in the exchange and hopping cases. However, due to the
$j$ dependence of $\Delta_j$, the condition defining the above
short-time limit is different in the two cases. In fact, the difference
is removed when the number of environmental modes becomes large, and the
effective coupling $\Lambda=\sqrt{\sum_k\lambda_k^2}$ increases
accordingly, so that
\begin{equation}
t\ll\frac{1}{\sqrt{|(\nu\mp\omega)^2 \pm \Lambda^2}|}
\underset{{\text{large-}}N}{~~~~~\longrightarrow~~~~~}t\ll\frac{1}{\Lambda}\,,
\label{e.short-t_large-N}
\end{equation}
which establishes a relation between the short-time constraint and some
large-$N$ condition that will be further discussed later on.

Overall, we have that the interaction (either exchange or hopping) of an
oscillator with a bosonic environment induces a dynamics that is amenable to a
description in terms of the interaction
with a fluctuating classical field if the
following conditions can be, at least approximately, met:
\begin{enumerate}
  \renewcommand{\theenumi}{(\emph{\roman{enumi}})}
\renewcommand{\labelenumi}{\theenumi}
  \item \label{c1} narrow environmental energy spectrum ($\omega_k \simeq \omega$ $\forall k$)
  \item \label{c2} short interacting times
  \item \label{c3} environment at thermal equilibrium.
\end{enumerate}

It is worth noticing that, if conditions \ref{c1}-\ref{c3} hold, the above
description in terms of classical fields is valid at all temperatures.

\section{Magnetic environment}
\label{s.magnetic}
We now consider the situation described by the
Hamiltonian~\eqref{e.H-spin_i}, i.e. that of a bosonic mode A
interacting linearly with a magnetic system S, made of $N$
spin-$\frac{1}{2}$ particles, each described by its respective Pauli
matrices $(\sigma_i ^x, \sigma_i ^y, \sigma_i ^z) \equiv \boldsymbol{\sigma}_i$.
As in Sec.~\ref{s.bosonic}, we consider both the exchange and the hopping case.
Setting 1) $g_{2i}=0$, with $g_i\equiv g_{1i}$ finite,
and 2) $g_{1i}=0$, with $g_i\equiv g_{2i}$ finite, $\forall i$, from
Eq.~\eqref{e.H-spin_i} we get
\begin{align}
\label{e.star-hamiltonianD}
H_1^\sms & = \nu a^{\dagger} a+
\sum_i f_i \sigma_i^z + \sum_i (g_i^*a \sigma_i^{+}
+ g_i a^{\dagger} \sigma_i^{-})\,, \\
\label{e.star-hamiltonianA}
H_2^\sms & = \nu a^{\dagger} a+
\sum_i f_i \sigma_i^z + \sum_i (g_i^*a^{\dagger} \sigma_i^{+}
+ g_i a \sigma_i^{-})\,,
\end{align}
where the superscript $S$ refers to the magnetic nature of the environment.
Setting $f_i=f$, $\forall i$, and
further choosing $f>0$, the EOM in the Heisenberg picture are
\begin{subequations}\label{e.EOM-DS}
\begin{align}
\mbox{Exchange:}\qquad\dot{a} & = i[H_1^\sms, a]=-i\nu a -i\sum_{i=1}^N
g_i\sigma_i ^-\,, \\
\dot{\sigma}_i^- & =i[H_1^\sms, \sigma_i^-]=-if\sigma_i ^-+iag_i^*2\sigma_i ^z \,,
\end{align}
\end{subequations}
\vspace*{-2\abovedisplayskip}
\begin{subequations}\label{e.EOM-AS}
  \begin{align}
  \mbox{Hopping:}\qquad\dot{a} & = i[H_2 ^\sms, a]=-i\nu a -i\sum_{i=1}^N g_i^*\sigma_i ^+\,,\\
  \dot{\sigma}_i^+ & =i[H_2 ^\sms, \sigma_i^+]=
  if\sigma_i^+-iag_i2\sigma_i ^z\,,
  \end{align}
\end{subequations}
where we have related the index of the Hamiltonians $H^S_{1,2}$
with the exchange and hopping cases, respectively.

Despite Eqs.~(\ref{e.EOM-DS})-(\ref{e.EOM-AS}) have the same form as Eqs.~(\ref{e.EOM-D})-(\ref{e.EOM-A}) of
the bosonic case, they cannot be solved exactly, due to the different
algebra of the spin operators.
However, restricting ourselves to physical
situations where the operator $S^z\equiv \sum_{i=1}^N\sigma_i^z$ can
be replaced by some reasonable expectation value
$\exval{S^z}\equiv \frac{N}{2}\exval{\sigma^z}\equiv-\frac{N}{2} {m}$,
(with ${m}>0$, due to $f$ being positive)
we can rewrite the above EOM in the form
\begin{subequations}\label{e.gl-EOM-DS}
  \begin{align}
  \mbox{Exchange:}\quad\qquad\dot{a} & =-i\nu a -i\Lambda^\sms\tilde S^- \,, \\
  \dot{\tilde{S}} ^-&=-if\tilde S^- -i \Lambda^\sms a
  \end{align}
\end{subequations}
\vspace*{-2\abovedisplayskip}
\begin{subequations}\label{e.gl-EOM-AS}
\begin{align}
\mbox{Hopping:}\quad\qquad\dot{a} & =-i\nu a -i\Lambda^\sms\tilde S^+ \,, \\
\dot{\tilde S}^+& =if\tilde S^+ +i\Lambda^\sms a\,,
\end{align}
\end{subequations}
with $g=\sqrt{\sum_{i=1}^N|g_i|^2}$,
$\Lambda^\sms=g\sqrt{2m}$, and
\begin{equation}
\tilde S^+=\frac{1}{\Lambda^\sms}\sum_{i=1} g_i\sigma_i^+~\,,~~
\tilde S^-=(\tilde S^+)^\dagger\,.
\label{e.global-op-unnormS}
\end{equation}
In fact, these equations can be derived from the
Hamiltonians
\begin{eqnarray}
&&\mbox{Exchange:}\quad \nu a^\dag a + f S^z +
\Lambda^\sms ( a \tilde{S}^+ + a^\dag \tilde{S}^-)\,,\;
\label{e.global-HDS}\\
&&\mbox{Hopping:~~}\quad \nu a^\dag a + f S^z +
\Lambda^\sms (a^\dag \tilde{S}^+ + a \tilde{S}^-)\,,
\label{e.global-HAS}
\end{eqnarray}
upon further assuming that the commutation relations
\begin{equation}
[\tilde{S}^+, \tilde{S}^-]=-1~\,,~~
[S^z,\tilde{S}^+]=\tilde{S}^+~\,,~~
[S^z,\tilde{S}^-]=-\tilde{S}^-\,,
\notag
\end{equation}
hold, meaning that the spin algebra is simplified into a bosonic one.

Notice that replacing the total spin operator $\sum_i \sigma_i^z$
with an expectation value $\exval{S^z}=\frac{N}{2}\exval{\sigma^z}$
we imply that the field $f$ selects the same
expectation value $\exval{\sigma^z}$ for every spin-$1/2$, in the
spirit of the usual random phase approximation.

Once linearized, the EOM (\ref{e.gl-EOM-DS})-(\ref{e.gl-EOM-AS})
can be solved as in the bosonic case, to get solutions formally
analogous to Eqs.~\eqref{e.EOM-solutions} for the operators $a$ and
$\tilde S_j$, with the replacement
$\mathcal{B}_j(t)\rightarrow\tilde S_j (t)$ with
$\tilde S_1=\tilde S^-$, $\tilde S_2=\tilde S^+$, and
$\omega \rightarrow f$ in the magnetic expressions corresponding
to Eqs.~\eqref{e.dDelta_j}.

Whatever follows Eq.~\eqref{e.EOM-solutions} in Sec.~\ref{s.bosonic} can
be easily retraced until the choice of the initial environmental state
$\rho_S$ appears into
\begin{align}
\rho_A(t) = \hbox{Tr}_S\left[ e^{-i H_j^\sms \!t} \rho_A
\otimes \rho_S\, e^{i H_j^\sms\! t} \right]\equiv {\cal E}^S_j(\rho_A)\,.
\label{e.spin-map}
\end{align}

Assuming that S is initially prepared in a state at
thermal equilibrium, we take
\begin{equation}
\rho_S = \frac{1}{1+n_\smt^\sms} \left(\frac{n_\smt^\sms}
{1+n_\smt^\sms}\right)^{\tilde S ^+ \tilde S ^-}\,,
\label{e.thermal-spin}
\end{equation}
with $n_\smt ^\sms\equiv \frac{N}{2} \left(1-m\right)$.

Despite the formal analogy with Eq. \eqref{e.rhoB-0}, it is important to
notice that the temperature-dependence of $n_\smt^\sms$, and hence that
of the dynamical map, is generally different from what we get in the
bosonic case, where the thermal number of photons is $n_\smt =
\left(\exp\{\frac\omega T\}-1\right)^{-1}$. We can, for example, suppose
that the magnetic environment thermalizes with the thermal bath so that
$\exval{S^z}=-\sign({f})S\mathit{B}_S (x)= -\frac{N}{2}\sign({f}) m$,
where $S=N/2$ and $\mathit{B}_S (x)=m$ is the Brillouin function
\begin{equation}
\mathit{B}_S(x)=\frac{2S+1}{2S}\coth\left(\frac{2S+1}{2S}x
\right)-\frac{1}{2S}\coth\left(\frac{x}{2S}\right)\,,
\label{e.Brillouin} \end{equation} with $x=S |f|/ T$. With this choice,
it is $n_\smt ^\sms\equiv S \left(1-\mathit{B}_S (x)\right)$ and the
dependence on T of the bosonic model is only recovered when
$T\rightarrow 0$, being $\mathit{B}_S(x)\rightarrow 1-e^{-x}$ the low
temperature limit of Eq.~(\ref{e.Brillouin}). Notice that, in order for
the above representation to stay meaningful in the large-$S$ limit,
temperature must scale as $T\sim S$ so as to guarantee a finite $x$;
performing such large-$S$ limit, the Brillouin function turns into the
Langevin one, $ \mathit{L}(x)=\coth(x)-\frac{1}{x}\,,
$ which is indeed the classical limit of
Eq.~(\ref{e.Brillouin}).

We observe that the approximations introduced for the spin system are
consistent with our aim of finding an effective classical description
for the environment: indeed, once the total spin is guaranteed a
constant value $S$, a classical-like behavior is expected for a
spin-system when $S\gg 1$~\cite{Lieb1973, Yaffe1982}, and
the bosonic expansion given by the Holstein-Primakoff transformation
can be safely truncated at its lowest order $S^+\sim
b^\dagger$ (if $f>0$, $b^\dagger$ being a generic bosonic creation operator)
 \cite{Radcliffe71}.

We can now write the initial state $\rho_{\rm A}\otimes \rho_S$ using
the Glauber formula as in Eq.~\eqref{e.glauber-initial},
with the spin displacement operator defined as $D_{\tilde S}(\gamma)
= \exp\{\gamma \tilde S ^+ - \gamma^*
\tilde S ^-\}$ due to the choice $f>0$, and
hence $\exval{\sigma^z}<0$
(had we taken $f<0$ it would be
$D_{\tilde S}(\gamma) = \exp\{\gamma \tilde S^- - \gamma^*
\tilde S ^+\}$).
Using the solutions of the EOM (\ref{e.gl-EOM-DS})-(\ref{e.gl-EOM-AS}),
one can write the evolution of displacement operators and proceed
as done in the previous section up to Eq.~(\ref{e.mapA}),
thus obtaining that the dynamical map in the
magnetic case does also correspond to a Gaussian noise channel.
With the additional requirement of a random phase approximation,
an effective Hamiltonian of the form of Eq.~\eqref{e.HAeff}
can hence be written again, allowing us to conclude that
the set of conditions sufficient to find an effective
classical description is the same as in the bosonic model, the only difference
being in the temperature dependence of the standard deviation $\sigma^2$,
due to the different definition of $n_\smt^\sms$ in the magnetic case.

\section{Large-$N$ environment: deriving the classical fields}
\label{s.large-N}
In this section we take a more abstract view on the
problem of what happens to the principal system A when its environment
becomes macroscopic. For the sake of clarity we will specifically refer
to the results presented in Secs.~\ref{s.bosonic}-\ref{s.magnetic} and,
in particular, to the model \eqref{e.bosonic-ex}.

Our aim is to understand whether the emergence of
an effective Hamiltonian
$H^{\rm eff}_{\rm A}(\zeta(t))$ as in Eq.~\eqref{e.HAeff} is a general
feature of
OQS with macroscopic environments. We also aim at further clarifying
the meaning of the conditions \ref{c1}-\ref{c3} given at the end of Sec. \ref{s.bosonic}, and the
reasons why they seem to be utterly necessary in order to
obtain an effective Hamiltonian description. Following suggestions
from Refs.~\cite{Yaffe1982,pc1,CalvaniCGV13,FotiCV16}, the main idea is
to show that the emergence of $H^{\rm eff}_{\rm A}(\zeta)$ is related to
the crossover from a quantum to a classical
environment, possibly observed when the number of components
becomes very large. In fact, were the environment described by a
classical theory, its effects on the system would naturally be
represented by the classical fields $\zeta$.
\par
Before introducing the general approach we are going to adopt, let us
briefly recall some useful notions.

A quantum description of a physical system, or {\it quantum theory}
$Q$ for
short, is based on the introduction of (1) a Hilbert space
${\cal{H}}$, (2) a Lie product $[\cdot,\cdot]$ that defines the
commutation rules between the operators on
${\cal{H}}$, and (3) a Hamiltonian $H$.
Traceclass operators on ${\cal{H}}$ that represent physical observables
usually make up a vector space: this space, together
with the above Lie product, is the Lie algebra $g$ of the theory.
The expectation values $\exval{O}\equiv\exval{\psi|O|\psi}\in{\mathbb{R}}$
of Hermitian
operators, are the (only) physical outputs of the theory,
i.e. the experimentally accessible properties of the system.

On the other hand, a classical description of a physical system, or {\it
classical theory} $C$ for short, is defined by (1) a phase space
${\cal{C}}$, (2) a Poisson bracket $\{\cdot,\cdot\}$, and (3) a
Hamiltonian $h(\zeta)$, with $\zeta$ representing the set of conjugate
variables of the classical phase-space ${\cal C}$. Real functions
$O(\zeta)$ are the (only) physical outputs of the theory, in the same
sense as above.

The problem of whether or not a system made by quantum particles can be
described by a classical theory has been extensively studied in the last
decades of the last century. Different approaches (see for instance
Sec.VII of Ref.~\cite{Yaffe1982} for a thorough discussion) all
showed that a large number $N$ of quantum constituents is a necessary
condition for a system to admit a classical description, but yet it is not a
sufficient one, as confirmed by the experimental
observation of macroscopic quantum states.
In fact, further conditions must be satisfied,
that crucially involve symmetry properties of the original quantum
theory, and its Lie algebra.
Specifically, in Ref.~\cite{Yaffe1982} it is demonstrated that the
$N\to\infty$ limit of a quantum theory $Q_N$, hereafter indicated by
$Q_{N\to\infty}$, is a classical theory $C$
if $Q_N$ exhibits a {\it global symmetry}. This latter requirement
means that it must exist a group of unitary operators, each acting
non-trivially on all of the $N$ constituents,
that leave the physical observables of the theory invariant
(see Ref.~\footnote{If
the theory describes $N$ spin-$1/2$ particles interacting via an
isotropic Heisenberg-like magnetic exchange, one such symmetry can be
that defined by operators that rotate the spin of each particle of the
same angle. Take instead $N$ particles whose interaction only depends on
their
distance: the symmetry might be that defined by the same spatial
translation of each particle. As for non-interacting, identical, but
yet distinguishable particles, a possible global symmetry is that
defined by the permutation operators.} for some examples).
Indeed, one such symmetry guarantees the existence of a
simpler theory $Q_k$ (with $k$ a real parameter defined by $N$) whose
$k\to 0$ limit, hereafter indicated by $Q_{k\to 0}$, is physically
equivalent to $Q_{N\to \infty}$,
by this meaning that each expectation value that stays finite in
the latter limit can be obtained as some
expectation value provided by $Q_{k\to 0}$.
On the other hand, $Q_{k\to 0}$, is also a well
defined classical theory $C$, with phase-space ${\cal C}$  and
classical hamiltonian $h(\zeta)$, that therefore provides an effective
classical description of the
original many-particles quantum system in its macroscopic limit, through the chain $Q_{N\to\infty}=Q_{k\to 0}=C$ (see Fig.~3 of
Ref.~\cite{Liuzzo-ScorpoCV15a} for a graphical depiction of the relation
between $Q_k$, $Q_N$, and $C$).
Details of the procedure for deriving the above classical theory are
given in Appendix, according to the results presented in
Ref.~\cite{Yaffe1982}, and recently used in the framework of OQS
\cite{Liuzzo-ScorpoCV15a,Liuzzo-ScorpoCV15b}: suffice it here to say
that $Q_N$ and $Q_k$ are related by
the fact that their respective Lie algebras, $g_N$ and $g_k$, are
representations of different dimensionality of the same abstract algebra $g$.

Let us now get back to our problem, specifically concentrating upon the
model described by the Hamiltonian \eqref{e.H-bosonic_k}. In order to be
used in the framework of OQS dynamics, the results mentioned above and
the procedure described in Appendix need being generalized, as we deal
with the quantum theory of a bipartite system where just one of the two
constituents, namely the environment, is intended to become macroscopic.
However, due to the linear structure of the interactions entering
Eq.~\eqref{e.H-bosonic_k}, the procedure can still be applied as
follows.

Keeping in mind that we have to deal with physically meaningful Lie
algebras, we first notice that the coupling terms in
\eqref{e.H-bosonic_k} can be written as
$a$($a^\dagger$) tensor-times
some sum over $k$ of operators acting on ${\cal{H}}_{\rm B}$ iff
either $\lambda_{1k}=\lambda_{2k}$ or
$\lambda_{1(2)k}=0$, for all $k$. Taking one or the other of the above
conditions true is quite equivalent, as far as the following
construction is concerned: for the sake of clarity, and at variance with
what done in Secs.~\ref{s.bosonic}-\ref{s.magnetic}, we
specifically choose $\lambda_{2k}=0$ and
set $\lambda_{k}\equiv\lambda_{1k}$ finite, for all $k$, meaning that
we explicitly consider the exchange case only. Further taking
$\omega_k=\omega~\forall k$, as done in
Secs.~\ref{s.bosonic}-\ref{s.magnetic}, we can
define the global operators
\begin{equation}
E \equiv\frac{1}{N}\sum_k^N b^\dagger_k b_k~~~{\rm and}~~~
B\equiv\frac{1}{\sqrt{N\Lambda^2}}\sum_k^N\lambda_k b_k~\,,
\label{e.global-op-norm}
\end{equation}
with $\Lambda^2\equiv\sum_k^N|\lambda_{k}|^2$ as in
Eq.~\eqref{e.global-op-unnorm}, and write
the Hamiltonian \eqref{e.H-bosonic_k} as
\begin{equation}
H=\nu a^\dagger a + N\left[\frac{\Lambda}{\sqrt{N}}(a^\dagger
B +aB^\dagger)
+\omega E \right]~;
\label{e.HN}
\end{equation}
the way $N$ enters Eqs.~(\ref{e.global-op-norm}-\ref{e.HN})
is designed to recognize $\frac{1}{N}$ as the parameter
to quantify quantumness of the environment B, and let all the
operators, no matter whether acting on A, B, or A+B, independent on
the number of environmental modes.

The operators (\ref{e.global-op-norm}), together with
the identity, are easily seen to generate a Heisenberg algebra on ${\cal
H}_{\rm B}$, being
\begin{equation}
[B,B^\dagger]=\frac{1}{N}~\,,~~[B,E]=\frac{1}{N}B~\,,
~~[B^{\dagger},E]=-\frac{1}{N}B^{\dagger}\,.
\label{e.globalcommutators}
\end{equation}
However, this cannot be regarded as the Lie algebra $g_N$ of some
environmental theory, due to the presence of non commuting operators
acting on $A$ in Eq.~\eqref{e.HN}, unless the $N\to\infty$ limit is taken, as
shown below.

Explicitly referring to the example given in Sec.~IV of
Ref.~\cite{Yaffe1982} and the strategy described in Appendix, we
introduce the set of antihermitian operators
\begin{equation}
\{L(\epsilon,\beta)\equiv iN(\epsilon E+\beta^* B +\beta
B^\dagger)\}\,,
\label{e.Lie-Algebra}
\end{equation}
where $\beta\in{\mathbb{C}}$, with
$|\beta|\propto\frac{1}{\sqrt{N}}$, while
the coefficients $\epsilon\in{\mathbb{R}}$ do not depend on $N$.
In the large-$N$ limit, where terms which are bilinear
in $\beta$ and $\beta^*$ can be neglected due to their dependence
on $N$, it is $[L_1,L_2]= L_3$, with $L_i\equiv L_i(\epsilon_i,\beta_i)$,
$\beta_3=i(\epsilon_1\beta_2-\epsilon_2\beta_1)$,
and $\epsilon_3=0$, meaning that the set \eqref{e.Lie-Algebra} is
a Lie Algebra. This is indeed the algebra $g_N$ whose recognition
represents the first step towards the large-$N$ limit of the quantum
theory that describes B.
It is easily checked that a possible representation $g_k$, of
the same abstract algebra represented by $g_N$, is given by the
$2\times 2$ matrices
\begin{equation}
\left\{\ell(\epsilon,\beta)\equiv i\left(
\begin{matrix}
0 & \beta^*  \\
0 & \epsilon
\end{matrix}
\right)\right\}\,,
\label{e.Liematrix}
\end{equation}
being $[\ell_1,\ell_2]= \ell_3$, with $\ell_i\equiv
\ell_i(\epsilon_i,\beta_i)$,
and $\beta_3,\epsilon_3$ as above.
We underline that the choice of a representation $g_k$
that contains only either $\beta$
or $\beta^*$ is also the simplest way to make the presence of
non-commuting operators on ${\cal H}_{\rm A}$ in the
Hamiltonian \eqref{e.HN}, harmless
as far as the following construction is concerned.
\par
The matrices $\ell(\epsilon,\beta)$ allow us to write
\begin{equation}
\left[L,\begin{pmatrix}1 \\ B\end{pmatrix}\right]\equiv
\begin{pmatrix} {[L,1]} \\ {[L,B]} \end{pmatrix}=
\begin{pmatrix}0 \\ -i(\epsilon B+\beta^*)\end{pmatrix}
\end{equation}
as
\begin{equation}
\left[L,\begin{pmatrix}1\\B\end{pmatrix}\right]=
\ell^\dagger\begin{pmatrix}1\\B\end{pmatrix}~,
\label{e.commB}
\end{equation}
with $\ell^\dagger\equiv(\ell^*)^t$, and, quite equivalently,
\begin{equation}
\left[L,\begin{pmatrix}1&B^\dagger\end{pmatrix}\right]=
\begin{pmatrix}1&B^\dagger\end{pmatrix}\ell\,.
\label{e.commBdagger}
\end{equation}
\par
Let us now consider the unitary operators
\begin{equation}
U(\epsilon,\beta)\equiv\exp\{L(\epsilon,\beta)\}~:
\label{e.Ueb}
\end{equation}
given that, for any pair of operators $O$ and $P$, it holds
\begin{equation}
e^{-P}Oe^{P}=\sum_n\frac{(-1)^n}{n!}
\underset{n~{\rm times}}{[P,[P,[...[P,}O]...]]]\,,
\label{e.nestedcomm}
\end{equation}
from Eqs.~\eqref{e.commB} and \eqref{e.commBdagger} it follows
\begin{equation}
U^{-1}\begin{pmatrix}1\\B\end{pmatrix}U=
u(\phi,\zeta)\begin{pmatrix}1\\B\end{pmatrix}\label{e.vectors-transform1}
\end{equation}
and
\begin{equation}
U^{-1}\begin{pmatrix}1&B^\dagger\end{pmatrix}U=
\begin{pmatrix}1&B^\dagger\end{pmatrix}u^\dagger(\phi,\zeta)\,,
\label{e.vectors-transform2}
\end{equation}
with
\begin{equation}
u(\phi,\zeta)\equiv
\left(\begin{matrix}
1&0\\
\zeta&\phi
\end{matrix}\right)\,,
\label{e.u}
\end{equation}
where
\begin{equation}
\phi=e^{i\epsilon}~~~{\rm and}~~~
\zeta=\frac{\beta}{\epsilon}\left( e^{i\epsilon}-1 \right)
\label{e.phipsi}
\end{equation}
are obtained by explicitly
summing the series in Eq.~\eqref{e.nestedcomm}.

The fact that the set \eqref{e.Lie-Algebra} is a Lie algebra in the
large-$N$ limit reflects upon the unitary operators $U(\phi,\zeta)$, in
that they form a group in the same limit.
In fact, this is just the Lie group corresponding to $g_k$,
sometimes dubbed {\it dynamical} \cite{ZhangFG90} or {\it
coherence \cite{Perelomov72} group}, that defines,
together with the arbitrary choice of a reference state
$\ket{0}\in{\cal{H}}_{\rm B}$, the Generalized Coherent States (GCS)
$\ket{u(\phi,\zeta)}\equiv U(\phi,\zeta)\ket{0}$ for the theory $Q_k$.
The reason why these states are so relevant, as further
commented upon in Appendix, is that the operators $B$
and $E$ are demonstrated \cite{Yaffe1982} to transform into
$B(u)\equiv\exval{u|B|u}/N$ and
$E(u)\equiv\exval{u|E|u}/N$,
respectively, as $N$ goes to infinity. Therefore, in order to find the
large-$N$ limit of the Hamiltonian \eqref{e.HN} we now only need to
evaluate $B(u)$ and $E(u)$, even without knowing the explicit form of
the GCS, to obtain $H^{\rm eff}_N(\zeta)$ from
\begin{eqnarray}
H\underset{N\to\infty}\rightarrow&~&
\nu a^\dagger a + N\left[ \frac{\Lambda}{\sqrt{N}}
\left(a^\dagger B(u)+aB^*(u)\right)+\omega E(u)\right]\nonumber\\
&~&\equiv H^{\rm eff}_N(\zeta)~,
\label{e.Hlim}
\end{eqnarray}
where the relation between $\ket{u}$ and $\zeta$ is
made explicit below.

To proceed accordingly, we choose the reference state for the
GCS: $\ket{0}=\Pi_k\ket{0}_k$, with $\ket{0}_k$ such that
$b_k\ket{0}_k=0$.
This implies, given the separable structure of the operators
$U(\phi,\zeta)$, that the states $\ket{u}$ are tensor products of
single-mode pure states.
As a consequence, it is $\exval{u|BB^\dagger|u}
=\exval{u|\sum_{k'k}b_{k'}b^\dagger_{k}|u}
=\exval{u|\sum_k b_kb^\dagger_k|u}=NE(u)$, which allows us to determine
$B(u)$ and $E(u)$ via
\begin{equation}
\bra{u}\begin{pmatrix}1\\B\end{pmatrix}\otimes
\begin{pmatrix}1&B^\dagger\end{pmatrix}\ket{u}= N
\left(\begin{matrix}
1&B^*(u)\\
B(u)&E(u)
\end{matrix}\right)\,,
\label{e.symbols}
\end{equation}
and finally obtain, by Eqs.~(\ref{e.vectors-transform1}-\ref{e.vectors-transform2}) and again
neglecting terms bilinear in $\beta$ and $\beta^*$,
\begin{align}
& \bra{0}
u(\phi,\zeta)\begin{pmatrix}1\\B\end{pmatrix}\otimes
\begin{pmatrix}1&B^\dagger\end{pmatrix}u^\dagger(\phi,\zeta)
\ket{0} = \notag \\
& = \braket{0 |
\left(\begin{matrix}
1&\zeta^*+\phi^*B^\dagger\\
\zeta+\phi B &~~\zeta\zeta^*+\zeta\phi^*B^\dagger+\zeta^*\phi
B+\phi\phi^*BB^\dagger\end{matrix}\right)
| 0} \notag \\
&= \left(\begin{matrix}1 & \zeta^* \\ \zeta& 1\end{matrix}\right)\,,
\end{align}
i.e. $E(u)=1/N$ and $B(u)=\zeta/N$.

The above
result implies that the original Hamiltonian
\eqref{e.HN} formally transforms, according to Eq.~\eqref{e.Hlim}, as
\begin{equation}
H\underset{N\to\infty}\longrightarrow H^{\rm eff}_{\rm A}(\zeta)=
(\nu a^\dagger a + \omega)+ \zeta^* a + \zeta a^\dagger\,,
\label{e.H-Bclassic}
\end{equation}
where we have rescaled $\zeta\rightarrow\zeta\Lambda /\sqrt{N}$ and
$(\zeta,\zeta^*)\in {\mathbb{R}^2}$ is generally proved
\cite{Yaffe1982} to
be any point of a classical phase-space ${\cal M}_{\rm B}$ with
canonical variables $q\equiv(\zeta+\zeta^*)/2$ and
$p\equiv(\zeta-\zeta^*)/(2i)$. Notice that $|\zeta|\propto
\Lambda/\sqrt{N}$, which is independent of $N$ by definition.

Once Eq.~\eqref{e.H-Bclassic} is obtained, we can maintain with
confidence that the Hamiltonian \eqref{e.H-bosonic_k}, originally acting
on A+B, formally transforms, as $N\to\infty$, into one that exclusively
acts on A: however, the presence of the classical field $\zeta$ is the
remnant of the underlying quantum interaction between A and the huge
number of elementary constituents of B, namely the bosonic modes
$\{b_k\}_{k=1}^N$. To this respect, notice that the Hilbert
space ${\cal H}_{\rm B}=\otimes_k {\cal H}_{{\rm b}_k}$
is replaced by a two-dimensional classical phase-space, ${\cal
M}_{\rm B}$, implying an impressive reduction of dynamical variables.
This reduction is the most striking consequence of the global symmetry
that the quantum theory for B must exhibit in order to flow
into a well defined classical theory when B is macroscopic.
In our case, although we did not explicitly used it, the
symmetry is that under permutation of the bosonic modes $b_k$, and that
is why we have set $\omega_k=\omega~\forall k$. In fact, one can easily
check that this is an essential condition for the very same definition
of global operators obeying commutation rules of the form
\eqref{e.globalcommutators}, which on their turn are necessary to
proceed to the definition of the Lie Algebra, and all the rest.
\par
At this point, we notice that $\omega_k=\omega~\forall k$
is just the ``narrow environmental energy-spectrum condition'' \ref{c1},
discussed at the end of Sec.~\ref{s.bosonic}. In fact,
it immediately strikes that the effective Hamiltonian in
Eq.~\eqref{e.H-Bclassic} has the same
structure of that in Eq.~\eqref{e.HAeff}, given that the latter refers
to an
interaction picture that hides the environmental frequency $\omega$. On
the other hand, it is somehow puzzling that time does not enter the
above construction, which leave us clueless, so far, concerning the
relation $\zeta\to\zeta(t)e^{-i\omega_\zeta t}$.
\par
Looking for the possible origin of a time-dependence in the classical
field $\zeta$, we reckon that the results of this section
imply the following. Suppose there exists another macroscopic system T
that is not coupled with A, and interacts with B in such a way that the
above global symmetry is preserved: the presence of T manifests itself
in terms of some parameter $\tau$ (think about time and/or temperature,
for instance) upon which $\zeta$ depends, according to the rule
$\zeta=\zeta(\tau)$ provided by the classical theory describing B+T.
This dependence can be safely imported into the effective
cdescription of A via $\zeta\to\zeta(\tau)$ in $H^{\rm eff}_{\rm A}(\zeta)$,
Eq.~\eqref{e.H-Bclassic}, as far as the direct interaction
between A and T can be neglected, at least on the time scales one is
interested in.

Finally, we notice that the detuning $\nu-\omega$ does not play any
role in this section, which brings us back to
Eq.~\eqref{e.short-t_large-N}
and the possible relation between the large-$N$ condition here enforced
and the short-time approximation previously adopted.
\section{Conclusions}
\label{s.conclusions}
In this paper, we have addressed the dynamics of a
bosonic system coupled to either a bosonic or a magnetic environment.
In particular, we have discussed the conditions under which
the dynamics of the system may be described in terms of the effective
interaction with a classical fluctuating field.
\par
Our results show that for both kinds of environments an effective,
time-dependent, Hamiltonian description may be obtained for short
interaction time and environments with a narrow energy spectrum at
thermal equilibrium. The corresponding dynamics is described by
a Gaussian noise channel independently of the kind of environment,
their magnetic or bosonic nature entering only the form of the noise
variance. As far as the energy spectrum is narrow, this effective
description is valid at all
temperatures and independently on the nature of the interaction between
the system and its environment.

Moreover, exploiting a general treatment based on the large-$N$ limit
of the environment, we have clarified the origin and the meaning of
the narrow-environmental-spectrum and short-time conditions.
In fact, we find that $\omega_k \simeq \omega$ $\forall k$ is needed for a global
symmetry to emerge and characterize the environment, which is a
necessary ingredient for the environment to be described by a small
number of macroscopic variables. On the other hand, the large
energy scale implied by whatever coupling with a macroscopic environment
limits any effective description to short times only.
\par
Overall, our results indicate that quantum environments may be described by
classical fields whenever global symmetries allows one to define environmental
operators that remain well defined when increasing the spatial size of the
environment. This is a quite general criterion that may serve
as a guideline for further analysis, e.g. for fermionic principal systems
and/or hybrid environments.

\begin{acknowledgments}
This work has been supported by UniMI
through the H2020 Transition Grant 15-6-3008000-625, and by
EU through the Collaborative Projects QuProCS (Grant Agreement No. 641277).
PV has worked in the framework of the Convenzione
Operativa between the Institute for Complex Systems of the
Italian National Research Council (CNR) and the Physics
and Astronomy Department of the University of Florence.
\end{acknowledgments}

\appendix*
\section{}
Consider a system made by $N$ elements which is described by
a quantum theory $Q_N$ that features a global symmetry, as defined
in Sec.~\ref{s.large-N} (we will equip quantities with the index $N$
to indicate their being relative to this $Q_N$ theory).

The procedure described in Ref. \cite{Yaffe1982} for
deriving the classical theory that formally represents
$Q_{N\to\infty}$ can be summarized as follows.

The first step  is
that of identifying $g_N$, exploiting the knowledge of the Hilbert
space ${\cal H}=\otimes_i^N{\cal H}_i$, the Lie product, and the
Hamiltonian $H_N$.\
As the Hamiltonian $H_N$ represents a physical observable, an effective
strategy to identify $g_N$ is that of writing $H_N$ as a linear
combination of operators, and see if they belong to some minimal set
that generates a representation $g_N$ of some abstract Lie algebra $g$.

The second step of the procedure is that of finding an
irreducible representation $g_k$ of $g_N$, which stands as the Lie
algebra
for $Q_k$ (notice that this most often implies that an explicit
expression for $H_k$ does also become available).
Here is where the existence of a global symmetry emerges as a necessary
ingredient, as it guarantees that the dimensionality of the
representation $g_k$ be significantly smaller than that of $g_N$.
In fact, the way $g_k$ can be most often identified is writing the
original
Hamiltonian as a linear combination of some global (i.e. acting non
trivially upon each subsystem) operators that
are invariant under the symmetry-operations, and generate a
representation of the same abstract algebra $g$ which is also
represented by $g_N$.

In the third step, generalized coherent states (GCS) for $Q_k$ come into
play: these are defined, according to the approach either of
Gilmore \cite{ZhangFG90} or, quite equivalently, of
Perelomov \cite{Perelomov72}, starting from the dynamical group of the
theory, which is nothing but the Lie group associated to $g_k$ by
the usual Lie correspondence \cite{Gilmore2012}, and is therefore
provided by the above second step.
GCS for $Q_k$, hereafter indicated by $\ket{u}\in{\cal H}_k$, enter the
procedure due to their being \cite{ZhangFG90} in
one-to-one correspondence with points $u$ on a manifold ${\cal M}_k$,
whose cotangent bundle is a classical phase-space ${\cal C}$.
In other terms, each GCS $\ket{u}$ of the theory $Q_k$
defines a point $u\in{\cal M}_k$ and a set of conjugate variables
$\zeta\in{\cal C}$.
In fact, it is demonstrated \cite{Yaffe1982} that $Q_{k\to 0}$
is a classical theory $C$,
with phase-space the above cotangent bundle ${\cal C}$, and hamiltonian
$h({\zeta})=\exval{u|H_k|u}/N$.

The last step of the procedure is that of deriving, possibly
without knowing the explicit form of the GCS, the exptectation
values $\exval{u|H_k|u}$, and finally obtain the effective classical
hamiltonian describing the original quantum system in the $N\to\infty$
limit.

The role of the parameters $N$ and $k$, which has been here
understood for the sake of a lighter narration, becomes evident when
explicitly employing the procedure, as in Sec.~\ref{s.large-N}, where it
is
$k=1/N$.
\bibliography{std-oqs.19.bib}
\end{document}